\documentclass[11pt]{article}
\usepackage{graphicx,amssymb,amsmath}
\usepackage{algorithmic}

\usepackage{epsfig, subfigure}

\newtheorem{Theorem}{Theorem}
\newtheorem{Definition}{Definition}

\newtheorem{Lemma}{Lemma}

\newtheorem{Corollary}[Lemma]{Corollary}

\newtheorem{Example}{Example}

\newenvironment{proof}{\noindent\textbf{Proof: }\ignorespaces}{}
\newcommand{\qed}{\hspace*{\fill}$\Box$\medskip}

\newcommand{\NR}{{\rm NR}}
\newcommand\eps\varepsilon

\newcommand{\plist}{{p_1,p_2,\ldots,p_n}}
\newcommand{\llist}{{\ell_1,\ell_2,\ldots,\ell_r}}
\newcommand{\call}{{\cal L}}
\newcommand{\calc}{{\cal C}}
\newcommand{\cale}{{\cal E}}
\newcommand{\calq}{{\cal Q}}
\newcommand{\cald}{{\cal D}}

\begin{document}

\title{More Efficient Algorithms and Analyses for Unequal Letter Cost Prefix-Free Coding}

\author{Mordecai  Golin\\ Hong Kong UST\\ {\em golin@cs.ust.hk}
\and
Li Jian\\ Fudan University\\ {\em lijian83@fudan.edu.cn}
}
%\author{
%\authorblockN{Mordecai Golin}
%\authorblockA{Dept of Computer
% Science \& Engineering\\
%Hong Kong UST\\
%Kowloon, Hong Kong\\
%golin@cs.ust.hk}
%\and
%\authorblockN{Jian Li}
%\authorblockA{Dept of Computer
%Science \& Engineering\\
%Fudan University\\
% Shanghai 200433,
% P.R.China.\\
%lijian83@fudan.edu.cn}
%}
\maketitle

%%%%%%%%%%%%%%%%%%%%%%%%%%%%%%%%%%%%%%%%%%%%%%%%%%%%%%%%%%%%%%%%%%%%
\begin{abstract}
There is a large literature devoted to
the problem of finding an optimal (min-cost) prefix-free code with an
unequal letter-cost encoding alphabet of size.  While there is no known
polynomial time algorithm for solving it optimally there are many good
heuristics that all provide additive errors to optimal.  The additive error
in these algorithms usually depends linearly upon the largest encoding
letter size.

This paper was motivated by the problem of finding optimal codes when
the encoding alphabet is infinite.  Because the largest letter cost
is  infinite, the previous analyses could give infinite error bounds.
We provide a new algorithm that works with infinite encoding alphabets.
When restricted to the finite alphabet case,  our algorithm often
provides better error bounds than the best previous ones known.


\end{abstract}

%%%%%%%%%%%%%%%%%%%%%%%%%%%%%%%%%%%%%%%%%%%%%%%%%%%%%%%%%%%%%%%%%%%%
\section{Introduction}

%%%%%%%%%%%%%%%%%%%%%%%%%%%%%%%%%%%%%%%%%%%%%%%%%%%%%%%%%%%%%%%%%%%%

\begin{figure*}[t]
\label{Fig:Code_Example}
$$
\begin{array}{|l|c|c|c|c|}
\hline
x & aaa & aab & ab  &b\\\hline
cost(x) & 3 & 5 & 4 &3 \\\hline
\end{array}
\hspace*{.4in}
\begin{array}{|l|c|c|c|c|}
\hline
x & aaa & aab & ab  &aaba\\\hline
cost(x) & 3 & 5 & 4 &6 \\\hline
\end{array}
$$
\caption
{In this example $\Sigma = \{a,b\}$.  The code on the left is
$\{aaa,aab,ab,b\}$ which is prefix free.  The code on the right is
$\{aaa,aab,ab,aaba\}$ which is not prefix-free
because $aab$ is a prefix
of $aaba$.  The second row of the tables contain the costs of the codewords when
$cost(a) =1$ and $cost(b)=3$.
}
\end{figure*}

\begin{figure*}[t]
\vspace*{.2in}
\centering
\includegraphics[width=\hsize]{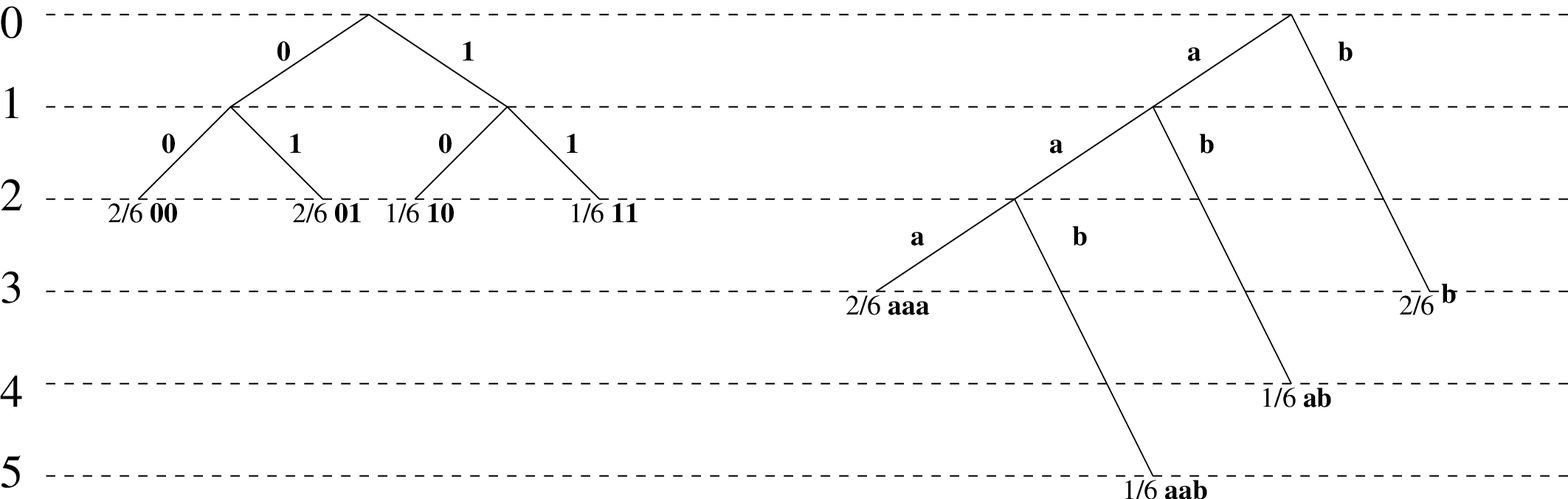}
\caption{Two min-cost prefix free codes for probabilities $2/6,2/6,1/6,1/6$ and their tree representations.
The code on the left is optimal for $c_1=c_2=1$ while the code on the right, the prefix-free code from Figure \ref{Fig:Code_Example}, is optimal for $c_1=1,c_2=3.$
}
\label{Fig:example}
\end{figure*}

Let $\Sigma = \{\sigma_1, \sigma_2.\ldots,\sigma_t\}$ be an {\em encoding alphabet}.   Word
$w \in \Sigma^*$ is a {\em prefix} of word $w' \in \Sigma^*$ if $w' = w u$ where $u \in \Sigma^*$ is
a non-empty word.  A {\em Code} over $\Sigma$ is a collection of words $C = \{w_1,\ldots,w_n\}$.  Code $C$ is {\em prefix-free}
if for all $i\not=j$ $w_i$ is not a prefix of $w_j.$  See Figure \ref{Fig:Code_Example}.

Let $cost(w)$ be the {\em length} or
number of characters in $w.$
Given a set of  associated probabilities
%or frequencies
\(p_1,  p_2, \ldots,  p_n \ge 0\), $\sum_i p_i =1,$  the cost of the code
is $Cost(C) = \sum_{i=1}^n cost(w_i) p_i$.
The {\em prefix coding} problem,
sometimes known as the {\em Huffman encoding} problem is to
find a prefix-free code over $\Sigma$ of minimum cost. This
problem is very well studied and has a well-known
$O(t n  \log n)$-time  greedy-algorithm  due to Huffman \cite{Huff-52}
($O(t n )$-time  if the $p_i$ are sorted in non-decreasing order).

{\em Alphabetic} coding is the same problem with the  additional constraint
that  the codewords must be chosen in increasing alphabetic
order (with respect to the words to be encoded). This corresponds,
for example, to the problem of constructing optimal (with respect to
average search time) search trees for items with the given access probabilities
or frequencies.  Such a code can be constructed
in $O(t n^3)$ time \cite{Itai-76}.

One well studied generalization of the problem  is to let
the encoding letters  have different costs.
That is, let $\sigma_i \in \Sigma$ have associated
cost $c_i.$ The cost of codeword $w = \sigma_{i_1} \sigma_{i_2} \ldots \sigma_{i_l}$ will
be $cost(w) = \sum_{k=1}^l c_{i_k}$, i.e., the sum of the costs of its
letters (rather than the length of the codeword) with the cost  of the code still being
defined as $Cost(C)=\sum_{i=1}^n cost(w_i) p_i$ with this new cost function.

The existing, large, literature
%As we will discuss immediately below there is  a very large literature
on the problem of finding a minimal-cost prefix free code when the $c_i$ are no longer equal, which will be surveyed below,
%All of it
assumes that $\Sigma$ is a finite alphabet,  i.e., that $t=|\Sigma|< \infty$.
The original motivation of this paper was to address the problem when $\Sigma$ is  {\em unbounded}.
which, as will briefly be described in  Section \ref{sec:motivation}
 models certain types of language restrictions on
prefix free codes and the imposition of different cost metrics on search trees. The tools
developed,  though,  turn out to provide improved approximation bounds for many of
the finite cases as well.
More specifically, it was known \cite{Krause-62,Mehlhorn-80}\footnote{Note that if $t=2$ with $c_1=c_2=1$
then $c=1$ and this reduces to the standard entropy lower bound for prefix-free coding.
Although the general lower bound is usually only
explicitly derived for finite $t$, Krause \cite{Krause-62} showed how to extend it to
infinite $t$ in cases where a positive root of $1 = \sum_{i=1}^\infty 2^{-c c_i}$ exists.} that
%{\em any} code for these probabilities must cost at least
$\frac 1 c H(p_1,\ldots, p_n) \le OPT$
where $H(p_1,\ldots,p_n) = - \sum_{i=1}^n p_i \log p_i$ is the
{\em entropy} of the distribution, $c$ is the unique positive root of the {\em characteristic
equation} $1 = \sum_{i=1}^t 2^{-c c_i}$ and $OPT$ is the minimum cost of any prefix free code
for those $p_i.$   Note that in this paper,  $\log x$ will always denote $\log_2 x.$

The known efficient algorithms create a  code $T$ that
satisfies
\begin{equation}
\label{eq:orig_bnd}
 C(T) \le  \frac 1 c H(p_1,\ldots, p_n) +  f(\calc)
 \end{equation}
where $C(T)$ is the cost of code $T$,  $ {\calc} = (c_1,  c_2,  \cdots, c_t)$
 and $f(\calc)$ is some function of the letter costs $\calc$, with
the actual value of $f(\calc)$ depending upon the particular algorithm.
Since $\frac 1 c H(p_1,\ldots, p_n) \le OPT$,   code $T$ has an {\em additive
error} at most $f(\calc)$ from $OPT.$
The  $f(\calc)$ corresponding to the different algorithms  shared an almost  linear dependence upon the value $c_t = \max(\calc),$ the largest letter cost.
They therefore can not be used for infinite $\calc.$  In this paper we present a new algorithmic variation
(all algorithms for this problem start with the same splitting procedure so they are all,  in some sense,  variations of each other) with a new analysis:
\begin{itemize}
\item (Theorems \ref{thm:beta} and \ref{thm:tbound})
For finite $\calc$ we derive new additive error bounds
$f(\calc)$ which in many cases, are much better than the
old ones.
\item (Lemma \ref{lem:Kbound})
If $\calc$ is infinite but $d_j = |\{m \mid  j \le c_m < j+1\}|$ is bounded,  then we can
still give a bound of type (\ref{eq:orig_bnd}).  For example, if $c_m= 1 + \lfloor \frac {m-1} 2 \rfloor$, 
i.e., exactly  two letters each of length $i, =1,2,3,\ldots$,  then we can show that
$f(\calc)\le 1 + \frac 3 {\log 3}$.
\item (Theorem \ref{thm:approx})
If $\calc$ is infinite but $d_i$ is unbounded then we can not provide a bound of type
(\ref{eq:orig_bnd}) but,  as long as
$\sum_{i=1}^{\infty} c_m 2^{-c c_m} < \infty$, we can show that
\begin{equation}
\label{eq:new_bnd}
 \forall \epsilon >0,\quad  C(T) \le (1+\epsilon)    \frac 1 c H(p_1,\ldots, p_n) +  f(\calc,\epsilon)
 \end{equation}
 where $f(\calc,\epsilon)$ is some constant based only on $\calc$ and $\epsilon$.
\end{itemize}

We now provide some more history and motivation.

For a simple example, refer to Figure
\ref{Fig:example}.  Both  codes  have minimum cost  for the frequencies
$(p_1,p_2,p_3,p_4) = \left(\frac 1 3 , \frac 1 3 , \frac 1 6 ,\frac 1 6\right)$ but under  different letter
costs.  The code   $\{00,01,10,11\}$
has  minimum cost for the standard Huffman problem in which
of $\Sigma = \{0,1\}$ and $c_1=c_2=1$,
i.e., the cost of a word is the number of bits it contains.
The code  $\{aaa, aab, ab, b\}$ has
minimum cost for the alphabet $\Sigma=\{a , b\}$
in which the length of an ``$a$'' is 1 and the length of a ``$b$'' is 3, i.e.,
${\calc} = (1,3).$

The unequal letter cost coding problem was originally  motivated by
coding problems  in which  different characters have
different transmission times or storage costs
\cite{Blach-54, Marcus-57, Karp-61, Stan-70, Varn-71}.
One example is the {telegraph channel}
\cite{gilb-69, gilb-95,Krause-62} in which $\Sigma=\{ \cdot , - \}$ and
$c_1 =1, c_2=2$, i.e.,
in which dashes are twice as long as dots.  Another
is the $(a,b)$ run-length-limited codes used in magnetic and optical storage
\cite{Imm-99,GR-98},
in which the codewords are binary and constrained
so that each {\bf 1} must be preceded by at least $a$, and at most
$b$, {\bf 0}'s.
(This example can be modeled by the  unequal-cost letter problem
by using an encoding alphabet of $r=b-a+1$ characters
$\{0^k 1 : k=a,a+1,\ldots,b\}$
with associated costs $\{c_i = a+i-1\}$.)

The unequal letter cost {\em alphabetic} coding problem arises in designing testing procedures
in which the time required by a test depends upon the outcome
of the test \cite[6.2.2, ex.\ 33]{Knuth-73}
and has also been studied under the names {\em dichotomous
  search} \cite{Hind-90} or the {\em leaky shower} problem \cite{kr-89}.

The literature contains many algorithms for the unequal-cost coding problem.
Blachman \cite{Blach-54}, Marcus \cite{Marcus-57},
and (much later) Gilbert \cite{gilb-95}
give heuristic constructions without analyses of the costs of the codes they produced.  Karp gave
the first algorithm yielding an exact solution (assuming the letter costs are integers);
Karp's algorithm transforms the problem into an integer
program and does not run in polynomial time \cite{Karp-61}.
Later exact algorithms  based on dynamic programming were given by
Golin and Rote \cite{GR-98}  for arbitrary $t$ and a slightly more efficient one by
Bradford et.\ al.~\cite{BGLR-02} for $t=2.$. These algorithms run in $n^{\theta(c_t)}$ time where
$c_t$ is the cost of the largest letter.
Despite the extensive literature,
there is no known polynomial-time algorithm for the
%general problem of Huffman coding with a variable-length alphabet,
generalized problem, nor is the problem known to be NP-hard.
Golin, Kenyon and Young \cite{GKY-02}  provide a polynomial time approximation scheme (PTAS).  Their
algorithm is mainly theoretical and not useful in practice.
Finally,  in contrast to the non-alphabetic case,
alphabetic coding has a polynomial-time algorithm $O(t n^3)$ time algorithm \cite{Itai-76}.

Karp's result was followed by many efficient algorithms
\cite{Krause-62, Csiszar-69, Cot-77, Mehlhorn-80, AM-80}.
As mentioned above, $\frac 1 c H(p_1,\ldots,p_n) \le OPT$;
almost\footnote{As mentioned by Mehlhorn  \cite{Mehlhorn-80}, the result
of Cot \cite{Cot-77} is a bit  different. It's a redundancy bound
and not clear how to efficiently implement  as an algorithm.  Also, the redundancy bound is in a very
different form involving taking the ratio of roots of multiple equations
that makes it difficult to compare to the others in the
literature.}
all of these algorithms produce codes of cost
at most $C(T) \le \frac 1 c H(p_1,\ldots,p_n) + f(\calc)$
and therefore give solutions
within an {\em additive error} of optimal.
An important observation is
that the additive error in these papers
$f({\calc})$ somehow incorporate the cost of the largest letter $c_t = \max(\calc)$.
Typical in this regard   is Mehlhorn's algorithm \cite{Mehlhorn-80} which provides  a bound of
\begin{equation}
\label{eq:Mbound}
c C(T) - H(p_1,\ldots,p_n) \le (1-p_1-p_n) + c c_t
\end{equation}
Thus, none of the algorithms described can be used to address  infinite alphabets with unbounded letter costs.

The algorithms all work by starting with the
probabilities in some given order,  grouping
consecutive probabilities together according to some rule, assigning the same initial codeword prefix
 to all of the probabilities in the same group and then recursing.
They therefore  actually create alphabetic codes. Another unstated assumption in those papers (related to their definition of alphabetic coding) is that the order of the $c_m$ is given and must be maintained.

In this paper we are only interested in the general
coding problem and not the alphabetic one and will therefore have freedom  to dictate the
original order in which the $p_i$ are given and the ordering of the $c_m.$
We will actually always assume that $p_1 \ge p_2 \ge p_2 \ge \cdots$ and $c_1 \le c_2 \le c_3 \le \cdots$.  These assumptions are the starting point that will permit us to derive better bounds.
Furthermore,  for simplicity,  we will always assume that $c_1=1.$  If not,  we can always force this by
uniformly scaling all of the $c_i.$

For further references on Huffman coding with unequal letter costs,
see Abrahams'  survey on source coding
%\cite[Section 2.7, {\em Coding with Unequal Cost Code Symbols: The Karp Problem}]{Abrahams-01},
\cite[Section 2.7]{Abrahams-01},
which contains a section on the problem.

%
%%%%%%%%%%%%%%%%%%%%%%%%%%%%%%%%%%%%%%%%%%%%%%%%%%%%%%%%%%%%%%%%%%%%
\section{Notations and definitions}
\label{sec:Defs}

There is a very standard correspondence between prefix-free codes over alphabet
$\Sigma$ and $|\Sigma|$-ary trees in which the $m^{\mbox{th}}$ child of node $v$ is labelled
with character $\sigma_m \in \Sigma.$  A path from the root in a tree to a leaf will
correspond to the word constructed by reading the edge labels while walking the path.
The tree $T$ corresponding to code $C=\{w_1,\ldots,w_n\}$ will be the tree containing
the paths corresponding to the respected words.  Note that the leaves in the tree will
then correspond
to codewords while internal nodes will correspond to prefixes of codewords.
See Figures \ref{Fig:example} and \ref{fig:Ex_1c}.
%\marginpar{Need a diagram}

Because this correspondence is 1-1 we will speak about codes and trees interchangeably,
with the cost of a tree being the cost of the associate code.

\begin{Definition}
Let $C$ be a prefix free code over $\Sigma$ and $T$ its associated tree.
$N_T$ will
denote the set of {\em internal nodes} of $T.$
\end{Definition}

%\begin{Definition}
%Let $p_1,p_2,\ldots,p_n \ge 0$ with $\sum_i p_i = 1.$   Their {\em entropy} is
%$$H(p_1,p_2,\ldots,p_n) = - \sum_{i=1}^n p_i \log p_i$$
%\end{Definition}

\begin{Definition}
Set $c$ to be the unique positive solution to $1 = \sum_{i=1}^t 2^{-c c_i}.$  Note that if $t < \infty$, then $c$ must exists while if $t=\infty$, $c$  might not exist.  We only define $c$ for the cases in which it exists. $c$ is sometimes called the
{\em root of  the characteristic equation} of the letter costs.
\end{Definition}

\begin{Definition}
Given letter costs $c_i$ and their associated characteristic root $c$, let $T$ be a code with those letter costs.
If  $p_1,p_2,\ldots,p_n \ge 0$ is a probability distribution then the {\em redundancy} of $T$ relative to the $p_i$ is
$$R(T;p_1,\ldots,p_n) = C(T) - \frac 1 c H(p_1,\ldots,p_n).$$
We\ will also define the {\em normalized redundancy} to be
$$\NR(T;p_1,\ldots,p_n) = c  R = c C(T) - H(p_1,\ldots,p_n).$$\
If the $p_i$ and $T$ are understood,  we will write $R(T)$ ($\NR(T)$) or even $R$ $(\NR$).
\end{Definition}
We note that many of the previous results in the literature, e.g., (\ref{eq:Mbound}) 
from \cite{Mehlhorn-80},
were stated in terms of $\NR.$  We will see later that this is a very natural measure for deriving bounds.
Also, note that by the lower bound previously mentioned, $C(T) \ge \frac 1 c H(p_1,\ldots,p_n)$ for all $T$ and $p_i$,
so  $R(T;p_1,\ldots,p_n)$ is a  good measure of absolute error.

%%%%%%%%%%%%%%%%%%%%%%%%%%%%%%%%%%%%%%%%%%%%%%%%%%%%%%%%%%%%%%%%%%%%
\section{Examples of Unequal-Cost Letters}
\label{sec:motivation}
%%%%%%%%%%%%%%%%%%%%%%%%%%%%%%%%%%%%%%%%%%%%%%%%%%%%%%%%%%%%%%%%%%%%

It is very easy to understand the unequal-cost letter problem
as modelling situations in which  different characters have
different transmission times or storage costs
\cite{Blach-54, Marcus-57, Karp-61, Stan-70, Varn-71}.  Such cases will all have
finite alphabets.  It is not a-priori as clear why infinite alphabets would be interesting.
We now discuss some motivation.

In what follows we will need some basic language
notation. A language $\call$ is just a set of words over alphabet $\Sigma.$ The
{\em concatenation} of languages $A$ and $B$ is $A  B = \{ab \mid a\in A, b \in B\}.$
The $i$-fold concatenation,  $\call^i$, is defined by
$\call^0 = \{\lambda\}$ (the language containing just the empty string),
$\call^1 = \call$ and $\call^i = \call \call^{i-1}.$
The {\em Kleene star} of $\call$, is
$\call = \bigcup_{i=0}^\infty \call^i$.

We start with cost vector
$\calc = \{1,2,3,\ldots,\}$ i.e, $\forall m>0, c_m =m.$
An early use  of this  problem was \cite{Patt69}.
The idea there was to construct a tree (not a  code)
in which the internal pointers to children were stored in a linked list.
Taking the $m^{\mbox{\small th}}$ pointer corresponds to using character $\sigma_m.$  The time that it
%takes to {\em find} the $i^{\mbox{\small th}}$ pointer is proportional to the location of the pointer in the
%LI
takes to {\em find} the $m^{\mbox{\small th}}$ pointer is proportional to the location of the pointer in the
list.  Thus (after normalizing time units) $c_m =m.$

We now consider a generalization of the problem of {\bf 1}-ended codes.  The problem of
finding min-cost
prefix free code with the additional restriction that all codewords end with a 1 was studied in
\cite{BY90,CSP94} with the  motivation of designing self-synchronizing codes.
One can model this
problem as follows.  Let $\call$ be a language.  In our problem,
$$\call = \{w \in \{0,1\}^* \mid \mbox{the last letter in $w$ is a {\bf 1}}\}.$$
We say that a code $C$ is in $\call$ if $C \subseteq \call.$
 The problem is to find a minimum cost code among all codes in $\call.$

Now suppose further that $\call$ has the special property that $\call = \calq^*$ where $\calq$ is itself a
prefix-free language.  Then every word in $\call$ can be uniquely decomposed as the concatenation of
words in $\calq$. If the decomposition of $w \in \call$ is
$w = q_1 q_2 \ldots q_r$ for $q_i \in \calq$ then  $cost(w) = \sum_{i=1}^r cost(q_i).$
We can therefore model the problem of finding a minimum cost code among all codes in $\call$ by first creating
an infinite alphabet $\Sigma_\calq = \{\sigma_q \mid q \in Q\}$  with associated
cost vector $C_\calq$ (in which  the
length of $\sigma_q$ is  $cost(q)$) and  then solving the minimal cost coding problem for $\Sigma_\calq$ with those
associated costs.
For the
example of {\bf 1}-ended codes we set
$\calq = \{1,01,001,0001,\ldots\}$ and thus have $\calc=\{1,2,3,\ldots,\}$ i.e, an infinite alphabet
with $c_m=1$ for all $m \ge 1.$

Now consider generalizing the problem as follows.  Suppose we are given an
unequal cost coding problem with {\em finite} alphabet $\Sigma=\{\sigma_1,\ldots,\sigma_t\}$
and associated cost vector $\calc = (c_1,\ldots,c_t).$  Now let $\Sigma' \subset \Sigma$ and
define
$$\call = \Sigma^* \Sigma'
=  \{w \in \Sigma^* \mid \mbox{the last letter in $w$ is in $\Sigma'$}\}.$$
Now note that $\call = D^*$ where $D = (\Sigma- \Sigma')^* \Sigma'$ is a prefix-free language.
We can therefore model the problem of finding a minimum cost code among all codes in $\call$ by
solving an unequal cost coding problem with alphabet $\Sigma_D$  and $\calc_D.$  The important
observation is that
$$d_j = |\{d \in \Sigma_D \mid cost(d) = j\}|,$$
the number of letters in $\Sigma_D$ of length $j$, satisfies a linear recurrence relation.
Bounding redundancies for these types of $\calc$ will be discussed in Section \ref{sec:Examples}, Case 4.

As an illustration, consider $\Sigma=\{1,2,3\}$ with $\calc=(1,1,2)$ and $\Sigma'=\{1\}$;
our problem is find minimal cost prefix free codes in which all words end with a $1.$
$\call = \{1,2,3\}^* \{1\} = D^*$, where $D = \{2,3\}^* \{1\}.$  The number of characters
in $\Sigma_D$ with length $j$ is
$$d_1 =1,\,
d_2 = 1,\,
d_3 = 2,\,
d_4 = 3,\,
d_5 = 5,\
$$
and, in general, $d_{i+2} = d_{i+1} + d_i$, so $d_i = F_i,$ the Fibonacci numbers.

We conclude with a very natural  $\call$ for which we do {\em not}
know how to analyze the redundancy.  In Section \ref{sec:Examples}, Case 5 we will discuss why this
is difficult.

Let $\call$ be the set of all ``balanced'' binary words\footnote{This also generalizes a problem from
\cite{LLY04} which provides heuristics for constructing a min-cost prefix-free
code in which the expected number of {\bf 0}'s equals the expected number of  {\bf 1}'s.},
i.e., all words which contain exactly
as many {\bf 0}'s as {\bf 1}'s.
Note that $\call = \cald^*$ where $\cald$ is the set of
all non-empty balanced words $w$ such that no prefix of $w$ is balanced.
Let
$$\ell_j = |\{w \in \call \mid cost(w) = j\}|,\quad
d_j = |\{w \in D \mid cost(w) = j\}|
$$
and set $L(z) = \sum_{n=0}^\infty \ell_n z^n$  and
$D(z) = \sum_{n=0}^\infty d_n z^n$ to be their associated generating functions.
If $\call = D^*$, then standard generating function rules, see e.g., \cite{SeFl96},
state that $L(z) = (1-D(z))^{-1}.$
Observe that $l_n=0$ if $n$ is odd and $l_n =  {{n} \choose {n/2}}$ for $n$ even,
so
$$
L(z) = \sum_{n=0}^\infty {{2n} \choose n} (z^2)^n = \frac 1 {\sqrt{1-4 z^2}}
$$
and
$$\sum_{n=1}^\infty d_j z^j = D(z) = 1 - \sqrt{1 - 4 z^2}.$$
This can then be solved to see that, for even $n>0$,  $d_n = 2 C_{n/2 - 1}$ where $C_i = \frac 1 {i+1} {{2i} \choose i}$
is the $i^{\mbox{\small th}}$ Catalan number. For $n=0$ or odd $n$,  $d_n = 0.$

%%%%%%%%%%%%%%%%%%%%%%%%%%%%%%%%%%%%%%%%%%%%%%%%%%%%%%%%%%%%%%%%%%%%
\section{The algorithm}
\label{sec:alg}
%%%%%%%%%%%%%%%%%%%%%%%%%%%%%%%%%%%%%%%%%%%%%%%%%%%%%%%%%%%%%%%%%%%%

\begin{figure}[t]

\centerline{
{\epsfig{file=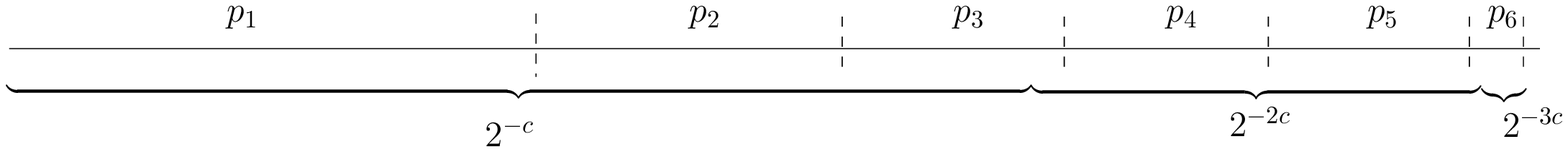, width=3.5in}}
\hspace*{.5in}
{\epsfig{file=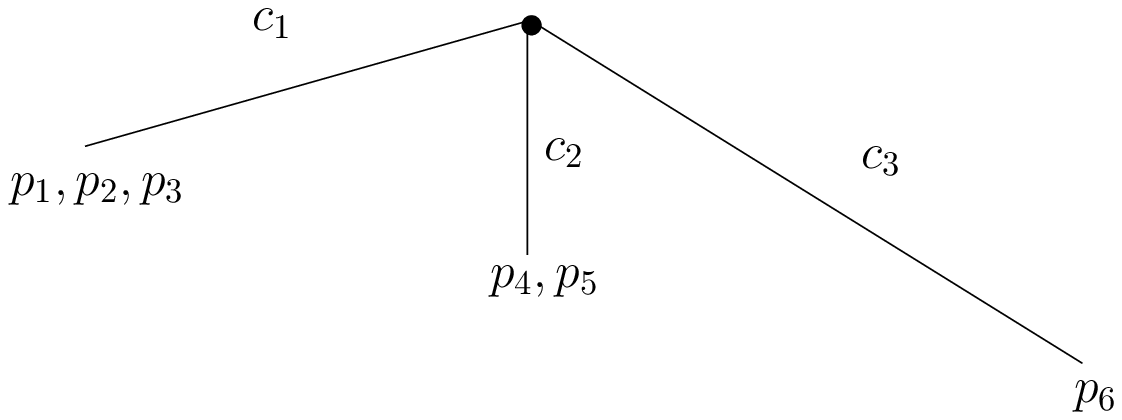, width=2in}}
}

\caption{The first splitting step for a case when $n=6$ $c_1=1$, $c_2=2,$ $c_3=3$ and the
associated preliminary tree.  This step groups $p_1,p_2,p_3$  as the first group,  $p_3,p_4$ as the second and $p_5$ by itself.  Note that we haven't yet formally explained yet {\em why} we've grouped the items this way.}
\label{fig:Ex_1a}

\vspace*{.4in}

\centerline{
{\epsfig{file=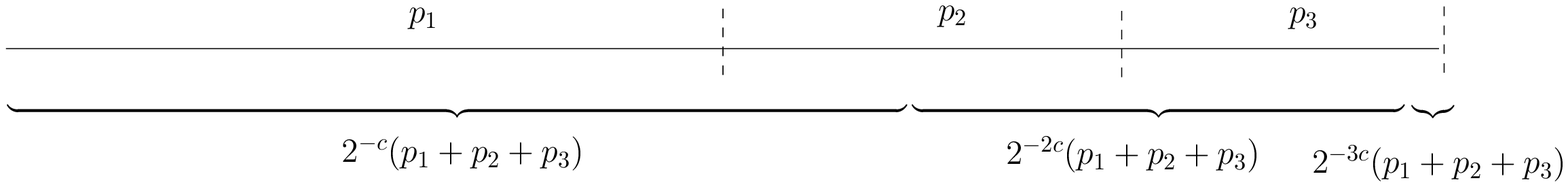, width=3.5in}}
\hspace*{.5in}
{\epsfig{file=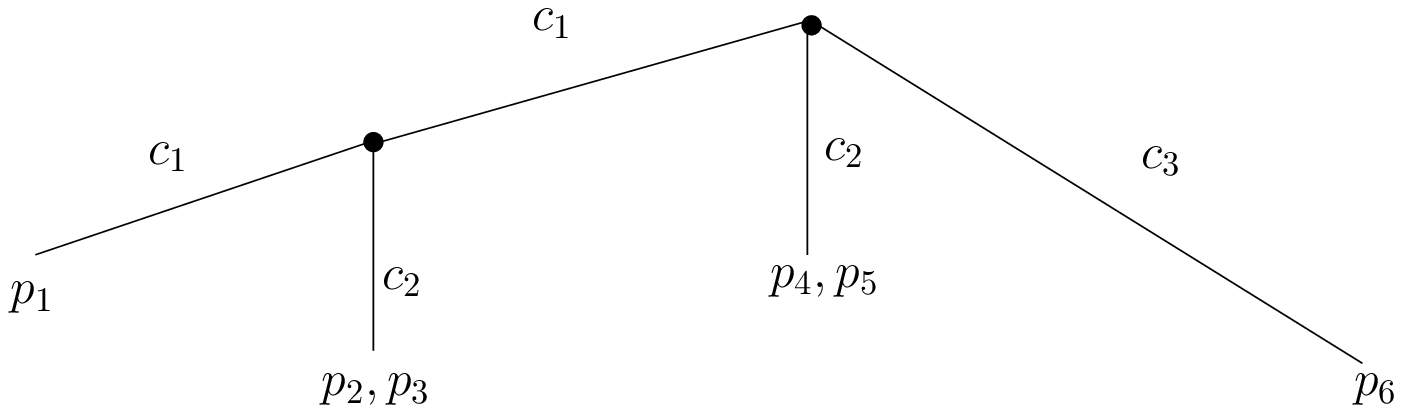, width=2in}}
}
\caption{In the second  split, $p_1$ is kept by itself and $p_2,p_3$ are grouped together.}
\label{fig:Ex_1b}

\vspace*{.4in}

\centerline{
\epsfig{file=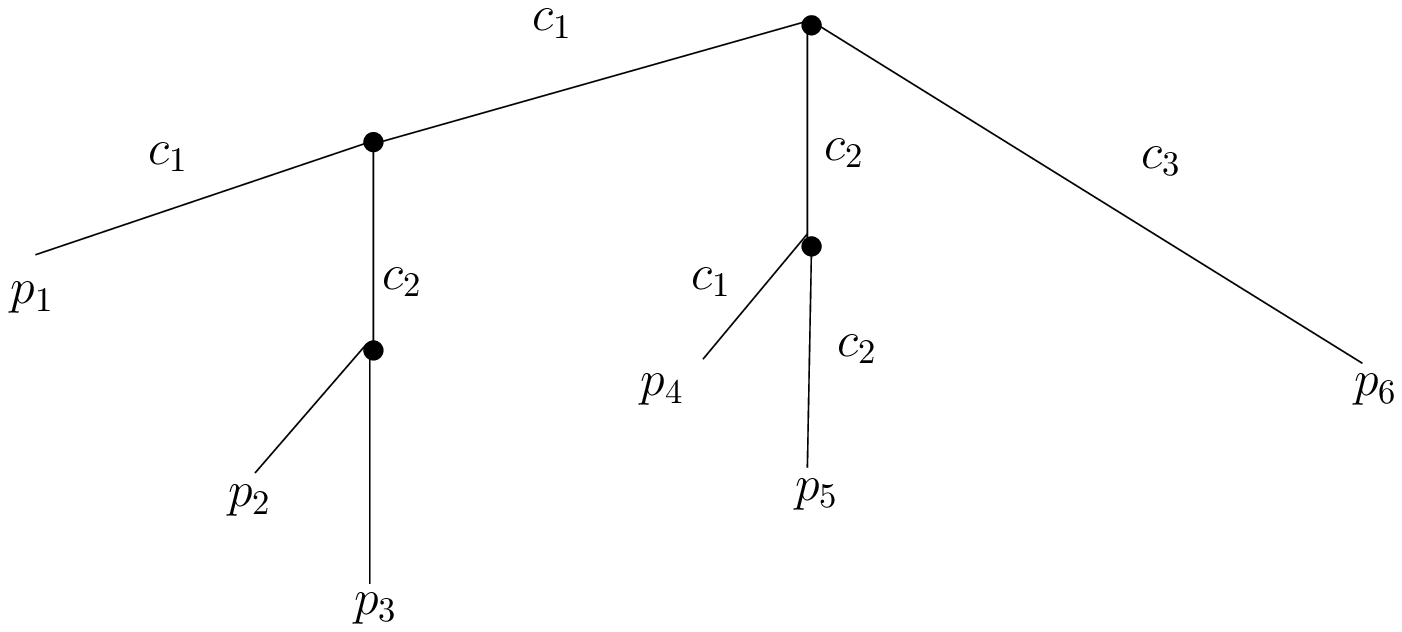, width=2in}
}
\caption{ After two more splits, the final coding tree is constructed.  The associated code is
$\{\sigma_1\sigma_1,\, \sigma_1\sigma_2\sigma_1,\, \sigma_1\sigma_2\sigma_2,\,
\sigma_2\sigma_1,\, \sigma_2\sigma_2,\, \sigma_3\}$
}
\label{fig:Ex_1c}
\end{figure}

All of the provably efficient heuristics for the problem, e.g., \cite{Krause-62, Csiszar-69, Cot-77, Mehlhorn-80, AM-80}, use the same basic approach,  which itself is a generalization of Shannon's original binary splitting algorithm \cite{shannon1948}. The idea is to create $t$ {\em bins}, where
bin $m$ has weight $2^{- c c_m}$ (so the sum of all bin weights is $1$). The algorithms then try to partition
 the probabilities into the bins;  bin $m$ will contain a set
of contiguous probabilities $p_{l_m},p_{l_m+1},\ldots,p_{r_m}$ whose sum
will have total weight ''close'' to $2^{-c c_m}.$  The algorithms fix the first letter of all the
codewords associated
with the  $p_k$ in  bin  $m$ to  be $\sigma_m.$  After fixing the first letter,
 the algorithms
then recurse, normalizing $p_{l_m},p_{l_m+1},\ldots,p_{r_m}$ to sum to $1$, taking them
as input and starting anew.
The various algorithms differ in how they group the probabilities and how they recurse.
See Figures \ref{fig:Ex_1a}, \ref{fig:Ex_1b} and  \ref{fig:Ex_1c} for an illustration of
this generic procedure.

Here we use a generalization of the version introduced in \cite{Mehlhorn-80}.
The algorithm first preprocesses the input and
calculates  all $P_k=p_1+p_2+\ldots+p_k$ ($P_0=0$)and $s_k = p_1+p_2+\ldots+ p_{k-1} + \frac {p_k} 2$.
Note that if we lay out the $p_i$ along the unit interval in order, then $s_k$ can
be seen as the {\em midpoint} of interval $p_i.$
It then partitions  the probabilities into  ranges,  and for each range
it constructs left and right boundaries $L_m,R_m$.  $p_k$ will be
assigned to bin $m$  if it ``falls'' into the ``range'' $[L_m,R_m)$.

If the interval $p_k$ falls into the range, i.e., $L_m \le P_{k-1} < P_k < R_m$
then $p_k$ should definitely be in bin $m$.  But what if $p_k$ spans two (or more) ranges, e.g.,
$L_m \le P_{k-1} < R_m < P_k$?  To which bin should $p_k$ be assigned?
 The choice made by \cite{Mehlhorn-80} is that $p_k$ is assigned to bin $m$ if $s_k=p_1+p_2+\ldots+p_k/2$ falls into   $[L_m,R_m)$,
  i.e., the midpoint of $p_k$ falls into the range.

%Although it is not often explicitly mentioned, the algorithms in \cite{Krause-62, Csiszar-69, Cot-77, Mehlhorn-80, %AM-80} assume that the {\em ordering} of the $p_i$ are given as part of the input. Therefore
%they are really solving the {\em alphabetic} coding problem and not the general one.  Also,  they implicitly
%assume that the $c_i$ appear in a given order as well.  While this makes sense for search trees, in the general coding %problem we have the flexibility to order the $c_i$ as we like.

%We therefore have the flexibility to require that  the $p_i$ are given in non-increasing order and the
%$c_i$ are given in non-decreasing order.  Both of these facts will be necessary for deriving our better %bounds.

Our procedure $CODE(l,r,U)$ will build a prefix-free code for $p_l,\ldots,p_r$ in which every code word starts with
prefix $U$.  To build the entire code we call  $CODE(1,n,\lambda)$, where $\lambda$ is the empty string.

\begin{figure}[p]

\centerline{
{\epsfig{file=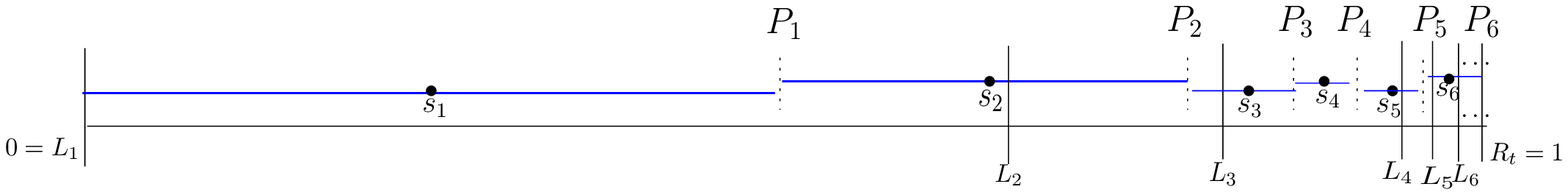, width=\hsize}}
}

\caption
%{The first step in our algorithm's  splitting procedure. $n=6$.  $L_i = \sum_{i=1}^i 2^{-c c_i}.$  Note that
%LI
{The first step in our algorithm's  splitting procedure. $n=6$.  $L_i = \sum_{m=1}^{i-1} 2^{-c c_m}.$  Note that
even though only the first 5 $L_i$ are shown,  there might be an infinite number of them (if $t = \infty$).
%Note too that, for $0 < i$,  $L_t = R_{t-1}.$
%LI
Note too that, for $0 < i$,  $L_i = R_{i-1}.$
}
\label{fig:Ex_2a}

\vspace*{.4in}

\centerline{
{\epsfig{file=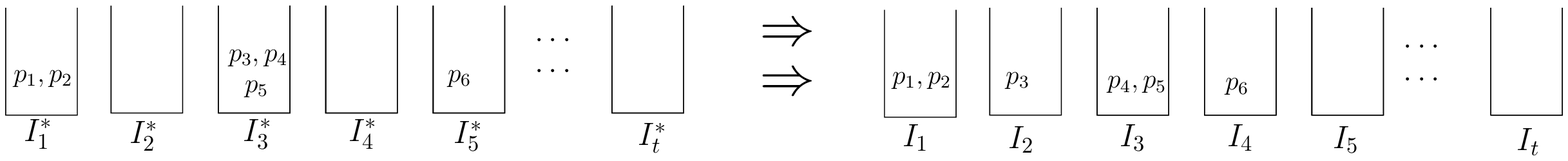, width=5in}}
}

\caption{The splitting procedure performed on the above example creates the bins $I^*_m$ on the left. The shifting procedure then creates the $I_m$ on the right.}
\label{fig:Ex_2b}

\vspace*{.4in}

%\begin{figure}[t]
\centerline{
{\epsfig{file=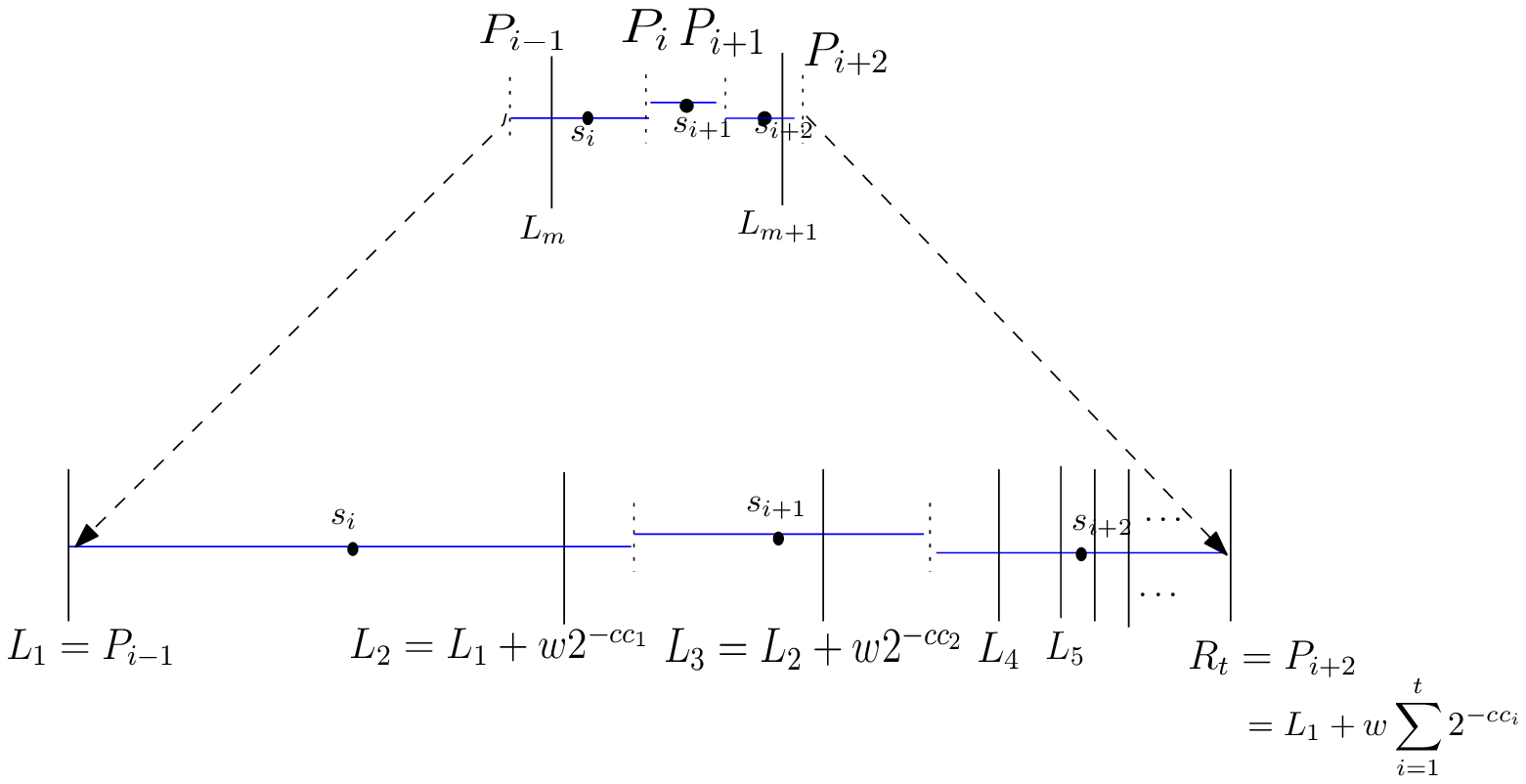, width=5in}}
}

\caption{An illustration of the recursive step of the algorithm.
$p_i,p_{i+1},p_{i+2}$ have been grouped together.  In the next splitting step,
the interval operated on
has length $w = p_i + p_{i+1} + p_{i+2}.$
}
\label{fig:rec}
%\end{figure}

\end{figure}

The  procedure  works as follows  (Figure \ref{fig:alg} gives pseudocode and Figures \ref{fig:Ex_2a},
\ref{fig:Ex_2b} and \ref{fig:rec}  illustrate the concepts):

\begin{figure*}[p]
\begin{tabbing}
$CODE(l,r,U)$;\\
\{{\em Constructs codewords $U_l,U_{l+1},\ldots,U_r$  for $p_l,p_{l+1},\ldots,p_r$.}\\
\hspace*{.02in} {\em  $U$ is previously constructed common prefix of $U_l,U_{l+1},\ldots,U_r$.\}}\\
{\bf If}  $l=r$\\
\hspace{.2in} \= {\bf then} codeword $U_l$ is set to be $U.$\\
{\bf else}    \> \quad {\em \{Distribute $p_i$s into initial bins $I^*_m$\}}\\
                            \> $L = P_{l-1}$; $R=P_r;$  $w = R-L$\\
%                           \> $\forall m,$ \= let $L_m = L + w\sum_{i=1}^{m-1} 2^{-c c_i}$ and $R_m = L_m + 2^{-c c_m}$.\\
%LI
                            \> $\forall m,$ \= let $L_m = L + w\sum_{i=1}^{m-1} 2^{-c c_i}$ and $R_m = L_m + w2^{-c c_m}$.\\
                            \>   \>set $I^*_m = \{k \mid  L_m \le s_k < R_m\}$\, \}\\[.05in]
                            \> \{\em Shift the bins to become final $I_m.$ Afterwards, \\
                            \>    \ {\em all bins $>M$ are empty, all bins $\le M$  non-empty}\\
                            \>  \ {\em and $\forall m \le M$, $I_m = \{l_m,\ldots r_m\}$\}}\\[0.1in]

                            \>  \{{\em shift left so there are no empty ``middle'' bins.}\}\\
                            \>  $M=0;$ $k=l$;\\
                            \>  {\bf while}  \= $k \le r$ {\bf do}\\
                            \>               \> $M=M+1;$\\
                            \>                           \> $l_M =k;$ $r_M = \max \Bigl(\{k\} \bigcup \{i>k \mid i \in I^*_M\}\Bigr)$;\\
                            \>                           \> $k = r_M+1;$\\[0.1in]

                            \>  \{\em If all $p_i$'s are in first bin,  shift $p_r$ to $2^{\mbox{nd}}$ bin \}\\
                            \> {\bf if} $r_1=r$ {\bf then}\\
                            \>   \> $M=2;$\\
                            \>   \> $r_1 = r-1$;  $l_2 = r_2 = r$;\\[0.1in]

                            \>  {\bf for} $m=1$ {\bf to} M {\bf do}\\
                            \>  \>      $CODE(l_m,r_m,U\sigma_m)$;
\end{tabbing}
\caption{Our algorithm. Note that the first step of creating the $I^*_m$ was written
to simplify the development of the analysis.  In practice,  it is not needed since
 $I^*_m$ is only used to
find $\max\{i > k \mid i \in I^*_m\}$ and this value can be calculated using binary search
at the time it is required.
}
\label{fig:alg}
\end{figure*}

Assume that we currently have a prefix of $U$ assigned to
$p_l,\ldots, p_r$. Let $v$ be node in the tree associated with $U.$ Let
$w(v)= \sum_{k=l}^r p_k.$

(i) If $l=r$ then word $U$ is assigned to $p_l$.  Correspondingly, $v$ is a leaf in the tree with
weight $w(v) = p_l.$

(ii) Otherwise let $L=P_l$ and $R=P_r$.  Split $R-L=w(v)$ into $t$ ranges\footnote{In the description, $t$
is permitted to be finite {\em } or infinite.}  as follows.
$$ \forall 1 \le m \le t,\quad
L_m = L + (R-L) \sum_{i=1}^{m-1} 2^{-c c_i},
\quad
R_m = L + (R-L) \sum_{i=1}^m 2^{-c c_i}.
$$
Insert $p_k$, $l \le k \le r$ in  bin $m$ if $s_k \in[L_m,R_m)$. Bin $m$ will thus contain the $p_k$ in
  $I^*_m(v) = \{k \mid  L_m \le s_k < R_m\}$.

We now shift the items $p_k$ {\em leftward}
in the bins as follows.  Walk through the bins from left to right.  If  the current
bin already contains some $p_k$, continue to the next bin.  If the current bin  is empty,  take the first $p_k$ that appears
in a bin to the right of the current one, shift $p_k$  into the current bin and walk to the next bin.  Stop when all $p_k$ have been seen.  Let
$I_m(v)$ denote  the items in the bins after this shifting.

Note that after performing this  shifting there is some $M(v)$ such that all bins $m \le M(v)$ are nonempty
and all bins $m > M(v)$ are empty.  Also notice that it is not necessary to actually construct the
$I^*_m(v)$ first.  We only did so because they will be useful in our later analysis.  We can more efficiently
construct the $I_m(v)$ from scratch  by walking from left  to  right, using a binary search each time, to find
the rightmost item that should be in the current bin. This will take  $O(M(v) \log (l-r))$ time in total.

We then check if all of the items are in $I_1(v)$.  If they are, we take $p_r$ and move it into $I_2(v)$
(and set $M(v) =2$).

Finally,  after creating the all of the $I_m(v)$ we let
$l_m=\min\{k\in I_m(v)\}$ and $r_m=\max\{k\in I_m(v)\}$
and recurse, for each $m< M(v)$  building $CODE(l_m,r_m,U\sigma_m)$

It is clear that the algorithm builds some prefix code with associated tree $T$.  As defined, let $N_T$ be the
set of internal nodes of $T.$  Since every internal node of $T$ has at least two children,
$\sum_{v \in N_T} M(v) \le 2n-1.$

The algorithm uses $O(1)$ time at each of its $n$ leaves and $O(M(v) \log n )$ time at node $v.$
Its total running time is thus bounded by
$$n + \sum_{v \in N_T}\log n  M(v) = O(n \log n)$$
with no dependence upon $t.$

For comparison, we point out the algorithm in  \cite{Mehlhorn-80} also starts  by first finding the
$I^*_m(v)$.  Since it  assumed $t < \infty$, its shifting stage was much simpler,  though.
It just shifted  $p_l$ into the first bin and $p_r$ into the $t^{\mbox{th}}$ bin (if they were
not already there).

We will now see that our modified shifting procedure not only permits a finite algorithm
for infinite encoding alphabets, but also  often provides a provably better approximation
for {\em finite} encoding alphabets.

%***********************
\section{Analysis}
%*************************

In the analysis we define $w^*_m(v) = \sum_{k \in I^*_m(v)} p_k$,
$w_m(v) = \sum_{k \in I_m(v)} p_k$.  Note that
$w(v) = \sum_{m=1}^t w^*_m(v) = \sum_{m=1}^t w_m(v) = \sum_{k=l}^r p_k$.

We first need three Lemmas  from \cite{Mehlhorn-80}.  The first was proven by recursion on the nodes of a tree, the second followed from the definition of the splitting procedure and the third from the second by some algebraic manipulations.

\begin{Lemma}\cite{Mehlhorn-80}
\label{lemma1}
Let $T$ be a code tree and $N_T$ be the set of all {\em internal} nodes of $T$. Then
\begin{enumerate}
    \item The cost $C(T)$ of the code tree T is
    $$C(T)=\sum_{v\in N_T}\sum_{m=1}^{t}c_m\cdot w_m(v)$$
    \item The entropy $H(p_1,p_2,\ldots,p_n)$ is
    $$H(p_1,p_2,\ldots,p_n)=\sum_{v\in N_T}w(v)\cdot H\left({w_1(v)\over w(v)},{w_2(v)\over w(v)},\ldots\right)$$
\end{enumerate}
\end{Lemma}

Lemma~\ref{lemma1} permits  expressing  the normalized redundancy of $T$ as
\begin{eqnarray*}
\label{f1}
 NR(T) &=&
c\cdot C(T)-H(p_1,p_2,\ldots,p_n)\\
&=&\sum_{v\in N_T} w(v)\left[\sum_{m=1}^{t}{w_m(v)\over w(v)}\left(\log2^{cc_m}+\log{w_m(v)\over w(v)}\right)\right].
\end{eqnarray*}

Set
$$
E(v,m)={w_m(v)\over w(v)}\left(\log2^{cc_m}+\log{w_m(v)\over w(v)}\right).
$$
Note that
$\NR(T)= \sum_{v\in N_T} w(v) \left(\sum_{m=1}^t E(v,m)\right).$
For convenience we will also define
\begin{eqnarray*}
E^*(v,m) &=& {w^*_m(v)\over w(v)}\left(\log2^{cc_m}+\log{w^*_m(v)\over w(v)}\right),\\
\NR^*(T)&=& \sum_{v\in N_T} w(v) \left(\sum_{m=1}^t E^*(v,m)\right)
\end{eqnarray*}
The analysis proceeds by bounding the values of $\NR^*(T)$ and $\NR(T) - \NR^*(T)$.

\begin{Lemma}
\label{lem:Mehl1}
\cite{Mehlhorn-80}\footnote{slightly rewritten for our notation}
({\em \small note: In this Lemma, the $p_i$ can be arbitrarily ordered.})\\
Consider any call $CODE(l,r,U)$ with $l < r.$ Let node $v$ correspond to the word $U$.
Let sets $I^*_1,I^*_2,\ldots$ be defined as in procedure CODE.
\begin{description}
    \item[a)] If $I^*_m=\emptyset$, then $w^*_m(v)=0$.
    \item[b)] If $I^*_m=\{e\}$. then $w^*_m(v)=p_e$.
%    \item[c)] $|I^*_m|\geq 2$ Let $e= \min I^*_m$ and $f = \max I^*_m$. \\
%LI
    \item[c)] If $|I^*_m|\geq 2$. Let $e= \min I^*_m$ and $f = \max I^*_m$. \\
    If $m=1$,   then $\frac {w^*_m(v)} {w(v)} \le 2^{-c c_1} + \frac {p_f} {2 w(v)} \le 2 \cdot 2^{-c c_1}$.  \\[.04in]
    If $m=t$ (note that this case requires $t < \infty$) then \\[.1in]
        \hspace*{.2in} $\frac {w^*_t(v)} {w(v)} \le 2^{-c c_t} + \frac {p_e} {2 w(v)} \le 2 \cdot 2^{-c c_t}$  \\[.04in]
         If $2 \le m < t,$ %\\[.1in]
        then $\frac {w^*_m(v)} {w(v)} \le 2^{-c c_m} + \frac {p_e + p_f} {2 w(v)} \le 2 \cdot 2^{-c c_m}$
%       \begin{itemize}
%       \item if $m=1$\\[.1in]
%       $\frac {w^*_m(v)} {w(v)} \le 2^{-c c_1} + \frac {p_f} {2 w(v)} \le 2 \cdot 2^{-c c_1}$  \\[.01in]
%        \item if $m=t$ (note that this case requires $t < \infty$)\\[.1in]
%       $\frac {w^*_t(v)} {w(v)} \le 2^{-c c_t} + \frac {p_e} {2 w(v)} \le 2 \cdot 2^{-c c_t}$  \\[.01in]
%       \item if $2 \le m < t,$\\[.1in]
%       $\frac {w^*_m(v)} {w(v)} \le 2^{-c c_m} + \frac {p_e + p_f} {2 w(v)} \le 2 \cdot 2^{-c c_m}$
%       \end{itemize}
\end{description}

\end{Lemma}

\begin{Lemma}\cite{Mehlhorn-80}
({\em \small note: In this Lemma, the $p_i$ can be arbitrarily ordered.})\\
In case (c) of Lemma \ref{lem:Mehl1}, $E^*(v,m) \le \frac{p_e + p_f} {w(v)}.\\$
Furthermore, if $m=1$ then $E^*(v,m) \le \frac{p_f} {w(v)}$,\\  while if $m =t$, then
$E^*(v,m) \le \frac{p_e} {w(v)}.$
\end{Lemma}

\begin{Corollary}
\label{cor:new}
If the $p_i$ are sorted in nondecreasing order then
in case (c) of Lemma \ref{lem:Mehl1},\\
\centerline{if $m=1$, $E^*(v,m) \le \frac{p_f} {w(v)},$ while
 if $m >1,$ then
$E^*(v,m) \le \frac{2 p_e} {w(v)}.$}
\end{Corollary}

\begin{Lemma}
\label{lem:combine}
$$\NR - \NR^* \le  c (c_2-c_1) \sum_{i \in A} p_i$$
where
$$A =  \{ i  \mid \mbox{$i$ is right shifted by the algorithm at some step}\}.$$
\end{Lemma}

\par\noindent
{\small \em Note: $p_1$ can never be right shifted, so   $\sum_{i \in A} p_i \le 1- p_1.$}

\medskip

\begin{proof}
Define
$$X(v) = \sum_{m=1}^t w(v) E(v,m)
\quad
\mbox{and}
\quad
X^*(v) = \sum_{m=1}^t w(v) E^*(v,m)
$$
Note that $\NR = \sum_{v \in N_T} X(v)$ and
$\NR^* = \sum_{v \in N_T} X^*(v)$
For each $v$ we will compare $X^*(v)$ and $X(v).$
If no shifts were performed while processing $v,$ then $X^*(v) = X(v)$ and there is nothing to do.
We now examine the two mutually exclusive cases  of performing left shifts or
performing a right  shift.

\medskip

\par\noindent \underline{Left shifts:}\\
Every step in our left-shifting procedure involves taking a probability out of
some bin $m$ and  and moving it into some currently empty bin  $r < m.$  Let
$w'_m(v) $ be the weight in bin $m$ before that shift and $p$ be the probability
of the item being shifted.  Note that the  original
weight of bin  $r$ was  $w'_r(v) = 0$ while after the shift, bin  $r$ will have weight $p$ and bin $m$
weight $w'_m(v) -p.$
We use the trivial fact
\begin{equation}
\label{eq:p+q}
\forall p,q >0,\quad
p \log p + q \log q \le (p+q) \log (p+q)
\end{equation}
Setting $q = w'_m(v) -p$ in (\ref{eq:p+q}) implies
$$p \log \frac {p} {w(v)} + (w'_m(v) -p) \log \frac { (w'_m(v) - p)} {w(v)}
\le w'_m(v) \log \frac {w'_m(v)} {w(v)}.$$
Furthermore,  the fact that the $c_i$ are nonincreasing implies
\begin{equation}
p \log 2^{c c_r} + (w'_m(v) -p) \log 2^{c c_m}
\le w'_m(v) \log 2^{c c_m}
\end{equation}
Combining the two last equations gives that
$$
p \left( \log 2^{c c_r} + \log \frac {p} {w(v)}\right)
+
(w'_m(v)-p) \left( \log 2^{c c_m} + \log \frac {w'_m(v)-p} {w(v)}\right)
$$
is
$$\le w'_m(v) \left( \log 2^{c c_m} + \log \frac {w'_m(v)} {w(v)}\right)
$$
Since moving from $X^*(v)$ to  $X(v)$ involves only operations in
which probabilities are shifted to the left into an empty bucket,
the analysis above implies that
$X(v) \le X^*(v).$

\medskip

\par\noindent \underline{Right shifts:}\\
Consider node $v$. Suppose that all of the probabilities in $v$ fall into $I^*_1$  with
$I^*_1= \{p_e,\ldots, p_f\}$ and $e \not=f.$
Since $p_f$ starts in bin 1,  $p_e$ must be totally contained in bin 1, so $p_e \le 2^{- c c_1} w(v)$.
The algorithm shifts $p_f$  to the right giving
 $I_1 = I_1 - \{p_f\}$ and $I_2=\{p_f\}$.
 The $p_i$ are nonincreasing so $p_f \le p_e.$
$$E(v,2) = \frac {p_f} {w(v)} \left(\log 2^{c c_2} + \log \frac {p_f} {w(v)}\right)
\le  \frac {p_f} {w(v)} (c c_2 - c c_1).$$
Also $E(v,1) \le E^*(v,1).$
Thus
\begin{eqnarray*}
w(v) \sum_{m=1}^t E(v,m)
&=& w(v) E(v,1) + w(v) E(v,2)\\
&\le& w(v) E^*(v,1) + p_f (c c_2 - c c_1)
\end{eqnarray*}

Once a $p_f$ is right-shifted it immediately becomes  a leaf and can never be right-shifted again.

Combining the analyses of left shifts and right shifts gives
$$\NR = \sum_{v \in N_T} X(v) \le \sum_{v \in N_T} X^*(v) + c(c_2-c_1)\sum_{i \in A} p_i = \NR^* + c(c_2-c_1) \sum_{i \in A} p_i.$$
\qed
\end{proof}

\begin{Lemma}
\label{lem:NR*_bound}
$$
\NR^* \le 2(1-p_1)  + \sum_{v \in N_T} \sum_{{1 \le m \le t} \atop {|I^*_m(v)| =1}} w(v) E^*(v,m).
%\NR^* \le 2(1-p_1) + (1 + \log \beta) \sum_{i \not\in A} p_i
% \log \beta  \sum_{v \in N_T} \sum_{{1 \le m \le t} \atop {|I^*_m(v)| =1}} w^*_m(v)
$$
\end{Lemma}

\begin{proof}
We evaluate $\NR^*$ by partitioning it into
\begin{equation}
\label{eq:NR_partition}
\NR^* = \sum_{v \in N_T} \sum_{{1 \le m \le t} \atop {|I^*_m(v)| \ge 2}} w(v) E^*(v,m)
+ \sum_{v \in N_T} \sum_{{1 \le m \le t} \atop {|I^*_m(v)| =1}} w(v) E^*(v,m).
\end{equation}
We use a generalization of an  amortization argument developed
  in \cite{Mehlhorn-80} to bound the first summand.
From Corollary \ref{cor:new} we know that if $|I^*_m(v)| \ge 2$
with $e= \min I^*_m$ and $f= \max I^*_m$ then
$w(v) E^*(v,m)$ is  at most  (a) $p_f$ or  (b) $2 p_e,$  depending upon whether
(a) $m=1,$  or (b) $m >1.$

Suppose that some $p_i$ appears as $2p_i$ in such a bound because  $i = \min I^*_m(v),$ i.e., case
(b).
Then, in all later  recursive steps of the algorithm $i$ will always be the
leftmost item in
bin 1 and will therefore  not be used in any later case (a) or (b) bound.

Now suppose that some $p_i$ appears  in such a bound because
$i = \max I^*_m(v),$ i.e., case (a).
Then in all later recursive steps of the algorithm, $i$ will always be the
rightmost item in the rightmost non-empty bin.  The only possibility for it to
be used in a later bound is  if becomes
the rightmost item in bin 1, i.e., all of the probabilities
%are in $I^*_1(v)$. In this case $p_i$ is used fo a second  case (a) bound.
%LI
are in $I^*_1(v)$. In this case, $p_i$ is used for a second  case (a) bound.
Note that if  this happens, then  $p_i$ is immediately right shifted, becomes  a leaf in
bin 2,  and is never used in any later recursion.

%Any given probability $p_i$ can therfore be used either once as
%LI
Any given probability $p_i$ can therefore be used either once as
a case (b) bound and contribute $2 p_i$ or twice as a case (b) bound and again contribute $2\cdot p_i.$
Furthermore, $p_1$ can never appear in a case (a) or (b) bound because, until
it becomes a leaf,  it can only be the leftmost item in bin 1.
 Thus
 \begin{equation}
\label{eq:new_bnd_2}
\sum_{v \in N_T} \sum_{{1 \le m \le t} \atop {|I^*_m(v)| \ge 2}} w(v) E^*(v,m) \le 2(1 - p_1).
\end{equation}
\qed
\end{proof}

{\em \small Note:
In Melhorn's original proof \cite{Mehlhorn-80}  the value corresponding to the RHS of
(\ref{eq:new_bnd_2}) was  $(1-p_1-p_n).$ This is because the shifting step of Mehlhorn's algorithm guaranteed that $|I^*_t(v)| \not=0$ and thus there was a symmetry between the analysis of leftmost and rightmost.  In our situation $t$ might be infinity so we can not assume that the rightmost non-empty bin is $t$ and  we get  $2(1-p_1)$ instead.}

\medskip
Combining this Lemma with Lemma \ref{lem:combine} gives
\begin{Corollary}
\label{cor:bound}
$$\NR \le 2(1-p_1) + c(c_2-c_1)\sum_{i \in A} p_i +
\sum_{v \in N_T} \sum_{{1 \le m \le t} \atop {|I^*_m(v)| =1}} w(v) E^*(v,m).$$
\end{Corollary}

We will now see different bounds on the last summand in the above expression. Section \ref{sec:Examples}
compares the results we get to previous ones for different classes of $\calc$.
 Before proceeding, we note that any $p_i$ can only appear  as
$I^*_m(v) = \{p_i\}$ for at most one $(m,v)$ pair.  Furthermore, if $p_i$ does appear in such a way,  then it can not
have been made a leaf by a previous right shift and thus  $p_i \not\in A.$

We start by noting that,  when $t \le \infty$ our bound is never worse than $1$ plus the old bound
of $(1-p_1-p_n) + c c_t$ stated in (\ref{eq:Mbound}).
\begin{Theorem}
\label{thm:first}
If $t < \infty$ then
$$\NR \le 2(1-p_1) + c c_t$$
\end{Theorem}
\begin{proof}
If $I^*_m(v) = \{p_i\}$ then $w(v) E^*(v,m) \le p_i c c_m$ so
%Since $w^*_m(v) \le w(v)$ we know $E^*_m(v) \le \frac {w^*_m(v)} {w(v)} (c c_m)$ so
$$\sum_{v \in N_T} \sum_{{1 \le m \le t} \atop {|I^*_m(v)| =1}} w(v) E^*(v,m)
\le \sum_{v \in N_T} \sum_{{1 \le m \le t} \atop {|I^*_m(v)| =1}} p_i c c_m
\le c c_t \sum_{i \not\in A} p_i.$$
The theorem  then follows from Corollary \ref{cor:bound}.
\qed
\end{proof}

For a tighter analysis we will need a better  bound for the case  $|I^*_m| =1$.
\begin{Lemma}
\label{lemma:wm}
(a) Let $v \in N_T.$  Suppose $i$ is such that $i \in I^*_m(v)$. then
$$
{p_i  \over w(v)}   \leq 2\cdot \sum_{j= m}^{t} {1\over 2^{c\cdot c_j}}
$$
(b) Further suppose there is some $m'>m$ such that $I^*_{m'} \not = \emptyset.$  Then
$$
{p_i  \over w(v)} \leq 2 \cdot \sum_{j=m}^{m'} {1\over 2^{c\cdot c_j}}
\leq 3 \cdot \sum_{j=m}^{m'-1} {1\over 2^{c\cdot c_j}}
$$
\end{Lemma}
\begin{proof}
Consider the call $CODE(l,r,U)$ at node $v$.  The fact that $i \in  I^*_m(v)$ implies
$L + \sum_{j=1}^{m-1} 2^{-c c_j} = L_m \le s_i$. To prove (a) just note that
$$  s_i + \frac {p_i} 2
= P_i \le P_r = R = L+ w(v) \sum_{i=1}^t 2^{-c c_i}.$$
So $\frac {p_i} 2  \le w(v) \sum_{j= m}^{t} {1\over 2^{c\cdot c_j}}$.

To prove part (b) let $i' \in I^*_{m'}.$  Then
$$  s_i + \frac {p_i} 2
= P_i \le s_{i'} < L+ w(v) \sum_{j=1}^{m'} 2^{-c c_j}.$$
So $\frac {p_i} 2  \le w(v) \sum_{j= m}^{m'} {1\over 2^{c\cdot c_j}}$.
The final inequality follows from the fact that $c_{m'-1} \le c_m.$
\qed
\end{proof}

\begin{Definition}
\label{def:beta}
Set $\beta_m = 2^{c c_m} \sum_{i=m}^t 2^{-c c_i}$ and $\beta = \sup \{\beta_m \mid 1 \le m \le t\}$
\end{Definition}

We can now prove our first improved bound:

\begin{Theorem} If $\beta < \infty$ then
\label{thm:beta}
%\label{lem:beta}
$$\NR \le 2(1-p_1) + \max\bigl(c(c_2-c_1), 1 + \log \beta\bigr)$$
\end{Theorem}
\begin{proof}
Note that using  Definition \ref{def:beta} and Lemma \ref{lemma:wm}(a)
we can bound the
last summand in Corollary
\ref{cor:bound} as
%\begin{eqnarray*}
% w(v) E^*(m,v) &=&
%w^*_m(v) \left(\log 2^{c c_m} + \log \frac {w^*_m(v)} {w(v)}\right)\\
%&\le& w^*_m(v) \log \left(2^{c c_m} 2 \sum_{i=m}^t 2^{-c c_i} \right)\\
%&\le&  w^*_m(v) (1 + \log \beta)
%\end{eqnarray*}
%LI
\begin{eqnarray*}
w(v) E^*(m,v)
&=&   w^*_m(v) \left(\log 2^{c c_m} + \log \frac {w^*_m(v)} {w(v)}\right)\\
&\le& w^*_m(v) \log \left(2^{c c_m} 2 \sum_{i=m}^t 2^{-c c_i} \right)\\
&\le& w^*_m(v) (1 + \log \beta)
\end{eqnarray*}
%we note that from Lemma \ref{lemma:wm}  and Definition \ref{def:beta}
If $w^*_m(v)=\{i\}$ then $i$ was not a leaf in any previous step and therefore could
not have been right shifted, so $i \not\in A.$ Thus
$$\sum_{v \in N_T} \sum_{{1 \le m \le t} \atop {|I^*_m(v)| =1}} w(v) E^*(v,m) \le (1+ \log \beta) \sum_{i \not\in A} p_i.$$
%Combining the bounds on the first and second summands proves the lemma.
\qed
\end{proof}

This immediately gives an improved bound for many finite cases because,
if $t < \infty$, then $\beta_m = 2^{c c_m} \sum_{i=m}^t 2^{-c c_i} \le t-m+1$ so $\beta \le t.$ Thus

\begin{Theorem}
\label{thm:tbound}
If $t$ is finite then
$$\NR \le 2(1-p_1) + \max\bigl(c(c_2-c_1), 1 + \log t \bigr)$$
\end{Theorem}

%In our results so far we never  made any assumption about the $c_i$ other than that $c_1 = 1.$ We now want to %assume that they are all integral and introduce some new notation.
\begin{Definition}
For all $j \ge 1$, set
$$d_j = |\{ i \mid  j \le c_i < j+1\}|.$$
\end{Definition}

This permits us to give another general bound that
also works for many infinite alphabets.
\begin{Lemma}
\label{lem:Kbound}
If  $d_j = O(1)$, then $NR = O(1)$.  In particular, if $\forall j,$  $d_j \le K$ then $\beta \le \frac {2^c K} {1-2^{-c}}$ so, from Theorem \ref{thm:beta},
%$$NR \le 2(1-p_1) + \max\left(c(c_2-c_1),\, 1 + c \log \left(\frac {K} {1-2^{-c}}\right)\right).$$
%LI
 $$NR \le 2(1-p_1) + \max\left(c(c_2-c_1),\, 1 + c + \log \left(\frac {K} {1-2^{-c}}\right)\right).$$
Furthermore, if all of the $c_i$ are integers, then $\beta \le \frac K {1-2^{-c}}$ and
$$NR \le 2(1-p_1) + \max\left(c(c_2-c_1),\, 1 + \log\left(\frac  K {1-2^{-c}}\right)\right).$$
\end{Lemma}

\begin{proof}
Since $c_1=1$ we must have $2^{-c} < 1.$
Thus, for all $m \ge 1$, if $\ell \le c_m < \ell+1$ then
\begin{eqnarray*}
\beta_m &=& 2^{c c_m} \sum_{i=m}^t 2^{-c c_m}\\
&\le& 2^{c (\ell+1)} \sum_{j=\ell}^\infty d_j 2^{-c j}\\
&\le& 2^c  K 2^{c \ell} \sum_{j=\ell}^\infty  2^{-c j} =\frac {2^cK} {1-2^{-c}}
\end{eqnarray*}
which is independent of $m$ and $\ell$.  The analysis when the $c_i$ are all integers is
similar. \qed
\end{proof}

For general infinite alphabets we are not able to derive a constant redundancy bound but we can prove
\begin{Theorem}
\label{thm:approx}
If $\calc$ is infinite and $\sum_{m=1}^{\infty} c_m 2^{-c c_m} < \infty$, then,
 for every $\epsilon >0$
\begin{equation}
\label{eq:new_bnd_3}
 R \le \epsilon  \frac 1 c H(p_1,\ldots, p_n) +  f(\calc,\epsilon)
 \end{equation}
 where $f(\calc,\epsilon)$ is some constant based only on $\calc$ and $\epsilon$. Note that this is equivalent
 to stating that
 $$C(T) \le (1+ \epsilon) OPT + f(\calc,\epsilon)$$
\end{Theorem}
\begin{proof}
We must bound the
$$\sum_{v \in N_T} \sum_{{1 \le m \le t} \atop {|I^*_m(v)| =1}} w(v) E^*(v,m) $$ term from the right hand side of Corollary \ref{cor:bound}.  Recall that $|I^*_m(v)| =1$ means that $\exists i$ such that
$I^*_m(v) =\{i\}$, i.e.,  $w^*_m(v) = p_i$ and  thus
%\begin{equation}
%\label{eq:sum_bnd}
$w(v) E^*(v,m) \le {p_i} c c_m.$
%\end{equation}

\medskip

Let $N_\epsilon$
be a value to be determined later and $m_{\epsilon}= \max \{m \mid c_m \le N_\epsilon\}$.  Since no
probability appears more than once in the sum we can write
$$\sum_{v \in N_T} \sum_{{1 \le m \le m_{\epsilon}} \atop {|I^*_m(v)| =1}} w(v) E^*(v,m)\le c N_\epsilon.
$$

To analyze the remaining cases, fix $v$.  Consider the set of indices
$$M_v = \{ m \mid  (m > m_{\epsilon}) \mbox{ and } |I^*_m(v) = 1|\}.$$
Sort these indices in increasing
order so that $M_v=\{m_1,m_2,\ldots,m_r\}$ for some $r$ with  $m_1 < m_2 < \cdots < m_r.$  Let $i_j$ be such that
$I^*_{m_j}(v) = \{p_{i_j}\}.$
Thus
%$$ \sum_{{m_{\epsilon}\le m } \atop {|I^*_m(v)| =1}} w(v) E^*(v,m)
%LI
$$ \sum_{{m_{\epsilon}< m } \atop {|I^*_m(v)| =1}} w(v) E^*(v,m)
= \sum_{j=1}^r w(v) E^*(v,m_j)
\le \sum_{j=1}^r  p_{i_j} c c_{m_j}
$$

Lemma  \ref{lemma:wm} and the fact that the $c_m$ are non-decreasing then gives
\begin{eqnarray*}
\sum_{j=1}^r  p_{i_j} c c_{m_j}
&\le& c w(v) \left[\sum_{j = 1}^{r-1} c_{m_j}\left( 3 \sum_{m=m_j}^{m_{j+1}-1} 2^{-c c_m}\right)
+ 2c_{m_r} \sum_{m=m_r}^\infty 2^{-c c_m}\right]\\
&\le& 3c w(v) \sum_{m=m_1}^\infty  c_m 2^{-c c_m}\\
&\le& 3c w(v) \sum_{m\ge m_{\epsilon}}^\infty c_m 2^{-c c_m}.
\end{eqnarray*}
We are given that $\sum_{m=1}^\infty c_m 2^{-c c_m}$ converges.  Thus $g(m_{\epsilon}) \downarrow 0$ as $m_{\epsilon} \rightarrow \infty$
where $g(x) = \sum_{m\ge x}^\infty c_m 2^{-c c_m}$.

Note that as $N_\epsilon$ increases,  $m_{\epsilon}$ increases.
Given $\epsilon$, we now  choose $N_\epsilon$ to be the smallest value such that $g(m_{\epsilon}) \le \frac \epsilon 6.$ Note that
$N_\epsilon$ is independent of $v.$

Combine the above bounds:
\begin{eqnarray*}
\sum_{v \in N_T} \sum_{{1 \le m \le t} \atop {|I^*_m(v)| =1}} w(v) E^*(v,m)
&=&
\sum_{v \in N_T} \sum_{{1 \le m \le m_{\epsilon}} \atop {|I^*_m(v)| =1}} w(v) E^*(v,m)
+
\sum_{v \in N_T} \sum_{{ m_{\epsilon} < m} \atop {|I^*_m(v)| =1}} w(v) E^*(v,m)\\
&\le&   c N_\epsilon + \sum_{v \in N_T}  \frac \epsilon  2 c w(v)
\end{eqnarray*}
Recall from Lemma \ref{lemma1} and the fact that $\forall m, c_m \ge 1,$
$$C(T) = \sum_{v\in N_T}\sum_{m=1}^{t}c_m\cdot w_m(v)
\ge \sum_{v\in N_T}\sum_{m=1}^{t} w_m(v)
= \sum_{v\in N_T} w(v).
$$
Thus,  we have just seen that
$$\sum_{v \in N_T} \sum_{{1 \le m \le t} \atop {|I^*_m(v)| =1}} w(v) E^*(v,m)
\le c N_\epsilon  +   \frac  \epsilon  2 c C(T).
$$

Plugging back into  Corollary \ref{cor:bound} gives
$$ c C(T) - H(p_1,\ldots,p_n) \le 2(1-p_1) + c (c_2-c_1) + c N_\epsilon  +   \frac \epsilon 2 c C(T)$$
which can be rewritten as
$$C(T) - \frac 1 {1- \frac \epsilon 2} \frac 1 c H(p_1,\ldots,p_n) \le \frac 1 {1- \frac \epsilon 2} \frac 1 c \left(2(1-p_1) + c (c_2-c_1) + c N_\epsilon\right)
$$
We may assume that $\epsilon \le 1/2,$ so $1+\epsilon \ge \frac 1 {1 - \frac \epsilon 2}$.  Thus
$$C(T) - (1+\epsilon) \frac 1 c H(p_1,\ldots,p_n) \le   f(\calc,\epsilon)$$
where
\begin{equation}
\label{eq:ebound}
f(\calc,\epsilon) = \frac 4 3 (\frac 2 c  +   (c_2-c_1) +   N_\epsilon).
\end{equation}
This can then be rewritten as
\begin{eqnarray*}
R = C(T) - \frac 1 c H(p_1,\ldots,p_n) &\le& \epsilon \frac 1 c H(p_1,\ldots,p_n)  +   f(\calc,\epsilon)\\
&\le& \epsilon OPT +  f(\calc,\epsilon)
\end{eqnarray*}
proving the Theorem.
\qed
\end{proof}

%***********************
\section{Examples}
\label{sec:Examples}
%*************************
We now examine some of the bounds derived in the last section and show how they compare to
the old bound of $(1-p_1-p_n) + c c_t$ stated in (\ref{eq:Mbound}).  In particular,  we show
that for large families of costs the old bounds go to infinity while the new ones
give uniformly constant bounds.

\medskip

\par\noindent\underline{Case 1:}\  $\calc_\alpha =(c_1,c_2,\ldots,c_{t-1},\alpha)$ with $\alpha\uparrow\infty.$ \\
We assume $t\ge 3$ and all of the $c_i,$ $i < t$, are fixed.
Let $c^{(\alpha)}$ be the root of the corresponding characteristic equation
$1= 2^{-c \alpha} + \sum_{i=1}^{t-1} c^{-c c_i}$.
Note that $c^{(\alpha)} \downarrow \bar c$
where $\bar c$ is the root of
$1=  \sum_{i=1}^{t-1} c^{-c c_i}$.  Let ($NR_{\alpha}$) $R_\alpha$  be the (normalized) redundancy corresponding
to $\calc_\alpha.$

For any fixed $\alpha$, the
old bound (\ref{eq:Mbound}) would give
$$NR_{\alpha} \le (1-p_1-p_n) + c^{(\alpha)}\alpha,
\quad
R_{\alpha} \le \frac {(1-p_1-p_n)} {c^{(\alpha)}}  + \alpha,
$$
the right hand sides of both of which tend to $\infty$ as
$\alpha$ increases.  Compare this to Theorem \ref{thm:tbound} which gives a uniform bound of
\begin{eqnarray*}
NR_{\alpha} &\le& 2(1-p_1) +  \max\bigl(c^{(\alpha)}(c_2-c_1), 1 + \log t \bigr)\\
&\le&  2(1-p_1) +  \max\left(c^{(c_{t-1})}(c_2-c_1), 1 + \log t \right)
\end{eqnarray*}
and
$$R_\alpha
\le \frac  {\NR_\alpha} {c^{(\alpha)}}
\le \frac {2(1-p_1) +  \max\left(c^{(c_{t-1})}(c_2-c_1), 1 + \log t \right)} {\bar c}.
$$

For concreteness,  we examine a special case of the above.
\begin{Example}
Let $t=3$ with $c_1=c_2=1$ and $c_3=\alpha \ge 1$.  The old bounds (\ref{eq:Mbound})  gives
an asymptotically infinite error as $\alpha \rightarrow \infty$.  The bound from
Theorem \ref{thm:tbound} is
$$
NR_{\alpha} \le  2(1-p_1) +  \max\bigl(c^{(\alpha)}(c_2-c_1),\, 1 + \log t \bigr) \le 3 + \log 3
$$
 independent of $\alpha.$
Since $c^{(\alpha)} \ge \bar c = 1$ we also get
$$R_\alpha = \frac {NR_\alpha} {c^{(\alpha)}}\le 3 +  \log 3.$$
%$$R_\alpha = \frac {NR_\alpha} {c^{(\alpha)}}\le 1 + \frac 1 3 \log 3.$$
%It is not difficult to derive that the positive root of
%$1 = 2^{-c} + 2^{-c} + 2^{-c \alpha}$ decreases monotonically to $1$ as
%$\alpha \rightarrow \infty$ with $c = 1 + O(2^{-\alpha})$ and  $ 2^{c} \le 3$.
%The old bound from **** was
%$$NR \le (1-p_1) + c \alpha,\
%\quad
%R \le \frac 1 c (1 - p_1) +  \alpha
%$$
%where the right hand side of both equations grows linearly to infinity in $\alpha$.

%To derive our new bound we note that for all $\alpha \ge  1$
%$$\beta_1 = 2^c (2 \cdot 2^{-c} + 2^{-c \alpha}) = 2^c \ge 2,\
%\quad
%\beta_2 \le \beta_1,
%\quad,
%\beta_3 = 2^{c \alpha} 2^{-c \alpha} =1
%$$
%so $\beta = 2^c$ and $\log \beta = c.$ Thus *** says that
%$$NR \le 2 (1-p_1) + 1 + c,
%\quad
%R \le \frac 2 c (1 - p_1) +  \frac  {1 + c} c
%$$
%which gives (very loose) uniform bounds of
%$$NR \le 3 + \log 3,
%\quad
%R \le 3 + \log 3.
%$$
\end{Example}

\medskip

\par\noindent\underline{Case 2:}\  A finite alphabet that approaches an infinite one.\\
Let $\calc$ be an infinite sequence of letter costs such that there exists a $K >0$ satisfying
for all $j,$  $d_j = |\{i \mid j \le c_i <j\}| \le K$.
  Let $c$ be the root of the characteristic equation $1 = \sum_{i=1}^\infty 2^{-c c_i}.$
Let $\Sigma^{(t)} = \{\sigma_1,\ldots,\sigma_t\}$ and its associated letter costs
be $\calc^{(t)} = \{c_1,\ldots,c_t\}$.   Let $c^{(t)}$ be the root of the corresponding
characteristic equation $1 = \sum_{i=1}^t 2^{-c c_i}$  and ($NR_t$) $R_t$ be the associated
(normalized) redundancy.
Note that $c^{(t)} \uparrow c$ as $t$ increases.

For any fixed $t$, the
old bound (\ref{eq:Mbound}) would be  $NR_{t} \le (1-p_1-p_n) + c^{(t)}c_t$ which goes to $\infty$ as
$t$ increases. Lemma \ref{lem:Kbound} tells us that
$$\beta^{(t)} = \max_{1 \le m \le t} 2^{c^{(t)} c_m} \sum_{i=1}^t 2^{c^{(t)} c_i}
\le  \frac {2^c K} {1-2^{-c^{(t)}}} \le
 \frac {2^c K} {1-2^{-c^{(2)}}}.
 $$
 so, from Theorem \ref{thm:beta} and the fact that $\forall t,$  $c^{(2)} \le c^{(t)} < c,$ we get
%$$NR_t \le 2(1-p_1) + \max\left(c(c_2-c_1),\, 1 + \frac K {1-2^{-c^{(2)}}}\right).$$
%LI
$$NR_t \le 2(1-p_1) + \max\left(c(c_2-c_1),\, 1 + c + \log\frac K {1-2^{-c^{(2)}}}\right).$$
Note that if all of the $c_m$ are integers,
%then Lemma \ref{lem:Kbound} will replace $2^c K$ by $K.$
%LI
then the additive factor $c$ will vanish.

\begin{Example}
Let $\calc=(1,2,3,\ldots).$ i.e., $c_m=m$.
The old bounds (\ref{eq:Mbound})  gives
an asymptotically infinite error as $\alpha \rightarrow \infty$.

For this case $c=1$ and  $K=1.$
$c^{(2)}$ is the root of the characteristic equation
$1 = 2^{-2} + 2^{-2c}$.
Solving gives $2^{-c^{(2)}} = \frac {\sqrt 5 -1} 2$ and $c^{(2)} = 1 - \log (\sqrt 5 -1) \approx 0.694\ldots.$
Plugging into our equations gives
\begin{eqnarray*}
NR_t &\le& 2(1-p_1) + \max\left(c(c_2-c_1),\, 1 + \log\left(\frac K {1-2^{-c^{(2)}}}\right)\right)\\
&=& 2 (1-p_1) + 1 + \log\left(\frac 2 {3 - \sqrt 5}\right) \le 4.388
\end{eqnarray*}
and
$$R_t = \frac {NR_t} {c^{(t)}} \le \frac {NR_t} {c^{(2)}} \le 6.232.$$
\end{Example}

\medskip
\par\noindent\underline{Case 3:}\  An infinite case when $d_j = O(1)$.\\
In this case just apply Lemma \ref{lem:Kbound} directly.

\begin{Example} Let $\calc$ contain $d$ copies each of $i=1,2,3,\ldots$, i.e.,
$c_m = 1 + \lfloor \frac {m-1} d \rfloor.$  Note that $K=d.$
If $d=1$, i.e., $c_m=m$, then
$c=K =1$ and
$$ R= \NR \le 2(1-p_1) + 2.$$
If $d>1$ then $A(x) = \sum_{m=1}^\infty c_m z^m = \frac {d z} {1-z}.$  The solution $\alpha$ to
$A(\alpha)=1$ is $\alpha = \frac 1 {d+1}$, so $c = - \log \alpha = \log (d+1).$
The lemma gives
$$\NR \le 2(1-p_1)  +\left(1 + \log\left(\frac K {1 - 2^{-c}}\right)\right) \le 3 +\log(d+1),\quad
R \le 1 + \frac {3} {\log(d+1)}.
$$
\end{Example}

\medskip
\par\noindent\underline{Case 4:}\   $d_j$ are integral and satisfy a linear recurrence relation.\\
In this case the generating function $A(z) = \sum_{j=1}^\infty d_j z^j = \sum_{m=1}^\infty z^{c_m}$ can be written as
$A(z) = \frac {P(z)} {Q(z)}$   where $P(z)$ and $Q(z)$ are relatively prime polynomials.  Let $\gamma$ be a smallest modulus root of $Q(z).$ If $\gamma$ is the unique root of that modulus (which happens in most
%interesting cases) then it is known that $d_j = \Theta(j^{d-1} \gamma^{-j}$
%LI
 interesting cases) then it is known that $d_j = \Theta(j^{d-1} \gamma^{-j})$
(which will also imply that $\gamma$ is positive real) where $d$ is the multiplicity of the root.  There must then exist some $\alpha < \gamma$ such that
$A(\alpha) =1.$ By definition $c = - \log \alpha.$ Furthermore,  since $\alpha < \gamma$ we must have that
%$\sum_{j=1}^\infty d_j j \alpha^j = \sum_{m=1}^\infty c_m z^{c_m}$ also converges, so
%LI
$\sum_{j=1}^\infty d_j j \alpha^j = \sum_{m=1}^\infty c_m \alpha^{c_m}$ also converges, so
Theorem \ref{thm:approx} applies.

Note that
$$ h(x) = \sum_{j=x}^\infty d_j j \alpha^j = O\left(\sum_{j=x}^\infty j^{d-1} j \left(\frac \alpha \gamma \right)^j\right)
= O\left(x^d \left(\frac \alpha \gamma \right)^x\right),
$$
implying
$$h^{-1}(\epsilon) = \log_{\gamma/\alpha} 1/\epsilon + O(\log\log 1/\epsilon)$$
where we define
%$$h^{-1} (\epsilon) = \max\{ x \mid h(x) > \epsilon,\, h(x-1) < \epsilon\}.$$
%LI
$$h^{-1} (\epsilon) = \max\{ x \mid h(x) \le \epsilon,\, h(x-1) > \epsilon\}.$$
Working through the proof of Theorem \ref{thm:approx} we find that when the $c_m$ are all integral,
\begin{eqnarray*}
\forall m',\quad
g(m') = \sum_{m \ge m'} c_m 2^{-c c_m}
            &=& \sum_{m \ge m'} c_m \alpha^{c_m} \\
            &\le& \sum_{j \ge c_{m'}} j d_j \alpha_j = h(c_{m'}).
\end{eqnarray*}
Recall that $m_\epsilon =\max\{m \mid c_m \le N_\epsilon\}$. Then
$g(m_\epsilon) \le h(N_\epsilon).$
%and, similarly,  for large enough $m'$,  $g(m') > h(c_{m'}-1).$
Since $g(m_\epsilon) \le \epsilon/6,$
$$N_\epsilon \le  h^{-1}(\epsilon/6) = \log_{\gamma/\alpha} 1/\epsilon + O(\log\log 1/\epsilon)$$
and thus our algorithm creates a code $T$ satisfying
\begin{equation}
\label{eq:lin_bnd}
C(T) - OPT \le \epsilon OPT +     \log_{\gamma/\alpha} 1/\epsilon + O(\log\log 1/\epsilon).
\end{equation}

\begin{Example}
Consider the case where $d_j = F_{j},$ the $j^{\mbox{th}}$ Fibonacci number, $F_1=1,$  $F_2=1,$  $F_3=2,$....
It's well known that $A(z) = \sum_{j=1}^\infty d_j z^j = \frac x {1 - x - x^2}$  and
$F_j = \frac {\phi^n - (1-\phi)^n} {\sqrt 5}$ where
$\phi = \frac {1 + \sqrt 5} 2$. Thus
$d_j = \frac {\gamma^{-j}} {\sqrt 5} + e_n$  where $\gamma = \phi^{-1}$ and $|e_n| < 1.$
 Solving $A(\alpha) = 1$ gives  $\alpha = \sqrt 2 -1 \approx .4142\ldots$
(and
$c = - \log \alpha = 1.2715\ldots$).
(\ref{eq:lin_bnd}) gives a bound on the cost of the redundancy of our code with
$\frac \gamma \alpha = \frac 2 {(1+\sqrt 5) (\sqrt 2 -1)} \approx 1.492\ldots.$
\end{Example}

\medskip
\par\noindent\underline{Case 5:}\ An example for which there is no known bound.\\
An interesting open question is how to bound the redundancy for
 the case of balanced words described at the end of Section \ref{sec:motivation}.
Recall that this had $d_j$ integral with $d_j=0$ for $j=0$ and odd $j$ and for even $j >0,$
$d_j = 2 C_{j/2 - 1}$ where $C_i = \frac 1 {i+1} {{2i} \choose i}$ is the $i^{\mbox{th}}$
{\em Catalan number}.
 It's well known that
$\sum_{j=0}^\infty C_j x^j = \frac 1 {2x} (1-\sqrt{1 - 4x})$ so
%$$A(x) = \sum_{m=1}^\infty c_m x^{c_m} = \sum_{j=1}^\infty d_j x^j =  1-\sqrt{1 - 4x^2}.$$  Solving for $A(\alpha) =1$ gives
%LI
$$A(x) = \sum_{m=1}^\infty  x^{c_m} = \sum_{j=1}^\infty d_j x^j =  1-\sqrt{1 - 4x^2}.$$  Solving for $A(\alpha) =1$ gives
$\alpha = \frac 1 2$ and $c = -\log \alpha = 1.$  On the other hand,
$$\sum_{m=1}^\infty c_m x^{c_m} = \sum_{j=1}^\infty j d_j x^j = 2 \sum_{j=1}^\infty  {{2 (j-1)} \choose {j-1}} (x^2)^j  = \frac {x^2} {\sqrt{1 - 4x^2}}$$
so  this sum {\em does not} converge when $x = 1/2.$  Thus,  we can not use Theorem \ref{thm:approx}
to bound the redundancy.  Some observation shows that this $\calc$ does not satisfy any
of our other theorems either.  It remains an open question as to how to construct a code
with ``small'' redundancy for this problem, i.e., a code with a constant additive approximation
or something similar to Theorem \ref{thm:approx}.

%***********************
\section{Conclusion and Open Questions}
\label{sec:conc}
%*************************
We have just seen $O(n \log n)$ time algorithms for constructing almost optimal prefix-free codes
for source letters with probabilities $p_1,\ldots,p_n$ when the costs of the letters of the
encoding alphabet are unequal values $\calc = \{c_1,c_2,\ldots\}.$  For many  finite encoding
alphabets, our algorithms have provably smaller redundancy (error) than previous algorithms given in
\cite{Krause-62, Csiszar-69, Mehlhorn-80, AM-80}.  Our algorithms also are the first that give provably bounded redundancy for some infinite alphabets.

There are still many open questions left.
The first arises by noting that, for the finite case, the previous algorithms were implicitly constructing
{\em alphabetic codes.}  Our proof explicitly uses the fact that we are only constructing general codes. It would
be interesting to examine whether it is possible to get better bounds for alphabetic codes (or to show that this is not possible).

Another open question concerns Theorem \ref{thm:approx} in which we showed that if
$\sum_{m=1}^{\infty} c_m 2^{-c c_m} < \infty$, then,
$$ \forall \epsilon >0,\quad C(T) - OPT  \le \epsilon  OPT  +  f(\calc,\epsilon).$$
Is it possible to improve this for some general $\calc$  to get a purely additive error rather than a
multiplicative one combined with an additive one?

Finally,  in Case 5 of the last section we gave a natural example for which
the root $c$ of $\sum_{i=1}^\infty 2^{- c c_m}=1$ exists but for which
$\sum_{m=1}^{\infty} c_m 2^{-c c_m} = \infty$ so that we can not apply Theorem \ref{thm:approx} and therefore
have no error bound.  It would be interesting to devise an analysis that would work for such cases as well.

\section*{Acknowledgments}
The first author's work was
partially supported by HK RGC Competitive
Research Grant 613105.

\bibliographystyle{plain}
\bibliography{Extended_Mehlhorn}

%***********************

\end{document}